**Ambient condition production of high quality reduced graphene oxide**

Sergey E. Svyakhovskiy*(1), Fedor S. Fedorov (2), Yuri A. Mankelevich (3), Pavel V. Dyakonov (3), Nikita V. Minaev (4), Sarkis A. Dagesyan (1), Konstantin I. Maslakov (5), Roman A. Khmelnitsky (6), Nikolay V. Suetin (3), Albert G. Nasibulin (2,7), Iskander Sh. Akhatov (8) and Stanislav A. Evlashin (8)

*sse@shg.ru

(1) Department of Physics, M. V. Lomonosov Moscow State University, 119991, Moscow, Russia

(2) Center for Photonics and Quantum Materials, Skolkovo Institute of Science and Technology, 143026, Moscow, Russia.

(3) D. V. Skobeltsyn Institute of Nuclear Physics, M. V. Lomonosov Moscow State University, 119991, Moscow, Russia

(4) Institute of Photonic Technologies, Research center "Crystallography and Photonics", RAS, 142190, Troitsk, Moscow, Russia.

(5) Department of Chemistry, M. V. Lomonosov Moscow State University, 119991, Moscow, Russia

(6) P. N. Lebedev Physical Institute, Russian Academy of Sciences, 119991 Moscow, Russia

(7) Department of Applied Physics, Aalto University, School of Science, P.O. Box 15100, FI-00076 Espoo, Finland

(8) Center for Design, Manufacturing and Materials, Skolkovo Institute of Science and Technology, 143026, Moscow, Russia

**Abstract**

Reduced graphene oxide becomes one of the most popular materials for applications in various optical, electronic and sensor devices. Even though many methods have been already reported for reduced graphene oxide synthesis, they usually rise issues related to their efficiency, quality and environmental impact. This work demonstrates a simple, environmentally friendly and effective method for reducing graphene oxide under ambient conditions using nanosecond infrared laser irradiation. As a result, a Raman band intensity ratio of I(G)/I(D) of 4.59 was achieved with an average crystallite size of ~90 nm. This graphene is of higher quality than what can be achieved with most of the existing methods. Additionally, the demonstrated reduction technique allows the selective reduction of graphene oxide and control the amount of functional groups on the surface of the material. Gas sensors fabricated according to the proposed technique efficiently detect $NO_2$, $NH_3$, and $H_2S$ with the sensitivity down to 10 ppm.

**1. Introduction**

There have been many approaches proposed to synthesize graphene oxide (GO)[1,2] and most of them are based on Hummers method introduced in 1958.[3] This method rapidly became widespread due to its simplicity and possibility to scale up the GO production. GO has been widely investigated and demonstrated in numerous applications,[4] e.g., photoluminescence,[5] transparent electrodes,[6] graphene paper,[7] water treatment,[8] cellular imaging and drug delivery.[9,10] In contrast to graphene, GO can be easily functionalized with various groups,[11–13] which allows to modify the properties of the material. However, there is a number of applications that simultaneously require the properties of GO with functional groups on its surface[14] and graphene with a high electrical conductivity.[15,16] To produce such kind of material GO can be thermally or chemically reduced.[17–21]

Recently, a few groups have reported selective laser-induced reduction of GO.[22] In particular, the formation of reduced graphene oxide (rGO) was examined by exposure to femtosecond laser irradiation.[23] Structural changes of the treated areas of GO were found to depend on the laser power.[23] However, the quality of the graphene in the reduced region was relatively low. The process of the GO reduction can be realized at various wavelengths and determined by pulse duration.[24] Nanosecond ultraviolet (UV) pulses are the most effective reduction method over a wide range of pulse durations from femtosecond to continuous irradiation. But the reduction was not complete, and the obtained rGO still possessed ~10% of its initial oxide groups according to X-ray photoelectron spectroscopy (XPS) measurements. Excimer laser reduction under an inert atmosphere (vacuum chamber) was demonstrated to produce rGO films with a high quality of a few percents of oxide groups and a very sharp 2D band in Raman spectroscopy.[25] Notably, a number of devices, including superconductors,[26,27] optical sensors,[28, 29] and water detectors,[30] have already been fabricated based on this approach.

The use of graphene in the fabrication of various types of sensors is of particular interest because such devices can detect even minor changes in the system configuration.[31-33] Single-layer graphene sensors can detect individual molecules[34] down to a detection limit of 1 ppb. However, the use of single- or multi-layered graphene films for device fabrication faces a number of technological issues such as graphene synthesis and transfer.[31] The use of rGO reduces the number of technological challenges and allows the fabrication process to be scaled up. rGO holds great potential as a foundation for gas detectors because of the large number of defects and functional groups are present on its surface. A number of theoretical papers have reported that the carbonyl and carboxyl groups,[35] as well as the defects,[36] in the rGO improve gas adsorption. Structural surface modifications can further improve gas adsorption, i.e., the effect of structural modifications on the adsorption energy was demonstrated for various types of gases.[37,38] Gas sensors with good performance based on GO and rGO have been prepared.[39-41]

In this paper, we introduce a simple ambient atmosphere laser reduction method to produce high-quality rGO with an oxygen content of less than 3%. Over the course of this study, we investigated more than 8000 sets of reduction conditions and determined the relationship between the parameters of the laser reduction and the quality of the obtained material. Such material is quite promising for the fabrication of highly sensitive gas sensors that detect $NO_2$, $H_2S$ and $NH_3$ at low concentrations.

## 2. Results

To determine the influence of laser irradiation conditions on the GO reduction, the following parameters were varied: laser pulse duration, pulse frequency, average laser power, and laser beam scanning speed. As a result, 8000 different sets of laser processing conditions were studied. The first step was to optically study the structures of the samples, and following selected samples were analyzed via Raman spectroscopy and scanning electron microscopy (SEM). The map of the studied samples is presented in **Figure 1**a. This photo shows the regions of GO reduced under different conditions. Some regions have almost no film visible (see Figure 1a area b), while other regions are fully covered with a film of rGO (Figure 1a areas c, d).

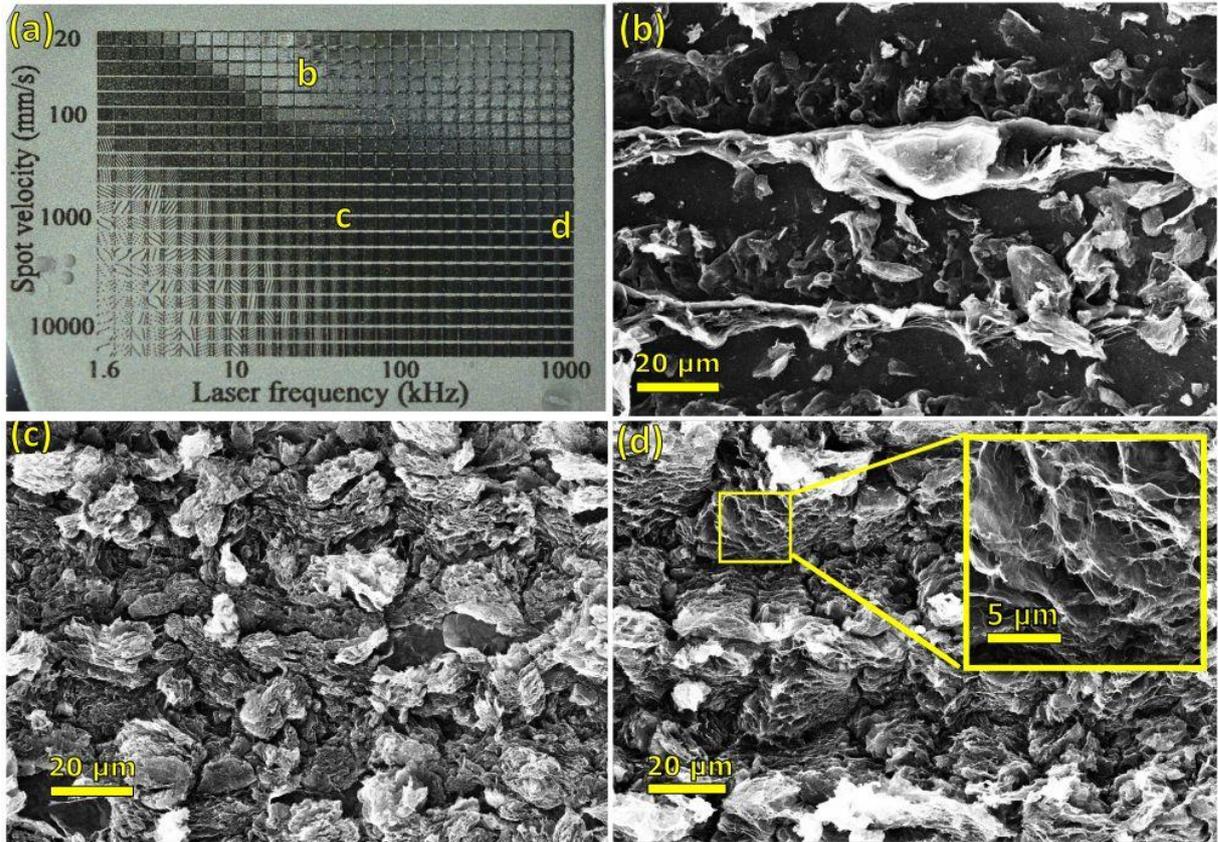

**Figure 1.** a) Photo of the surface after laser treatment with different laser frequencies and velocities of laser spot. b)-d) SEM images showing the structures of the different regions of the sample.

The white curve in Figure 1a indicates the optimal laser reduction conditions, which are concentrated in a limited range of energy fluences (EFs). In our regimes, the EF is proportional to the frequency (f) and the energy of the pulse ($E_{pulse}$) and inversely proportional to the velocity of the laser spot (v) and the beam diameter (D):

$$EF \approx \frac{f \times E_{pulse}}{v \times D}. \tag{1}$$

The white curve (Figure 1a) corresponds to nearly constant fluences due to constant f/v and $E_{pulse}$ at frequencies <20 kHz and due to constant v and f×$E_{pulse}$ at higher frequencies. The value of the EF is the major determining factor in the outcome of the GO reduction. Even though the regions shown in Figure 1a-d look alike to the naked eye and in the SEM micrographs, their Raman spectra are very different (see Figure S1 in the Supporting Information). **Figure 2**a-b shows the Raman spectra of the GO and rGO samples that are shown in Figure 1d. The I(G)/I(D) ratio is 0.97 for GO and 4.59 for rGO, and the I(2D)/I(G) ratio is 0.50 for rGO. Using the obtained I(G)/I(D) ratio and an empirical expression we can estimate the graphene crystallite size, which is approximately 18 nm for GO and 88 nm for rGO.[42] This estimated size is indicative of an rGO of considerably higher quality than what is indicated by the crystallite sizes of the rGO obtained by laser reduction in a vacuum.[25] Figure S1 shows more detailed information on the qualitative Raman studies on the rGO obtained under different reduction conditions. Samples with the best I(G)/I(D) ratio (see Figure 1d) were chosen for the XPS studies. The XPS results show that the GO film contains ~28% oxygen and ~70% carbon, while that of rGO has only approximately 3.5% oxygen and ~ 95.3% carbon (see Figure 2d). Table S1 and Figure S2 b-c provide more detailed data on the XPS studies of the products of various reduction conditions.

As shown in formula (1) the reduction of GO is controlled by the absorption of energy per unit surface area, i.e., the heating temperature. Bellow, we will estimate the laser heating and show the temperature distribution across the films.

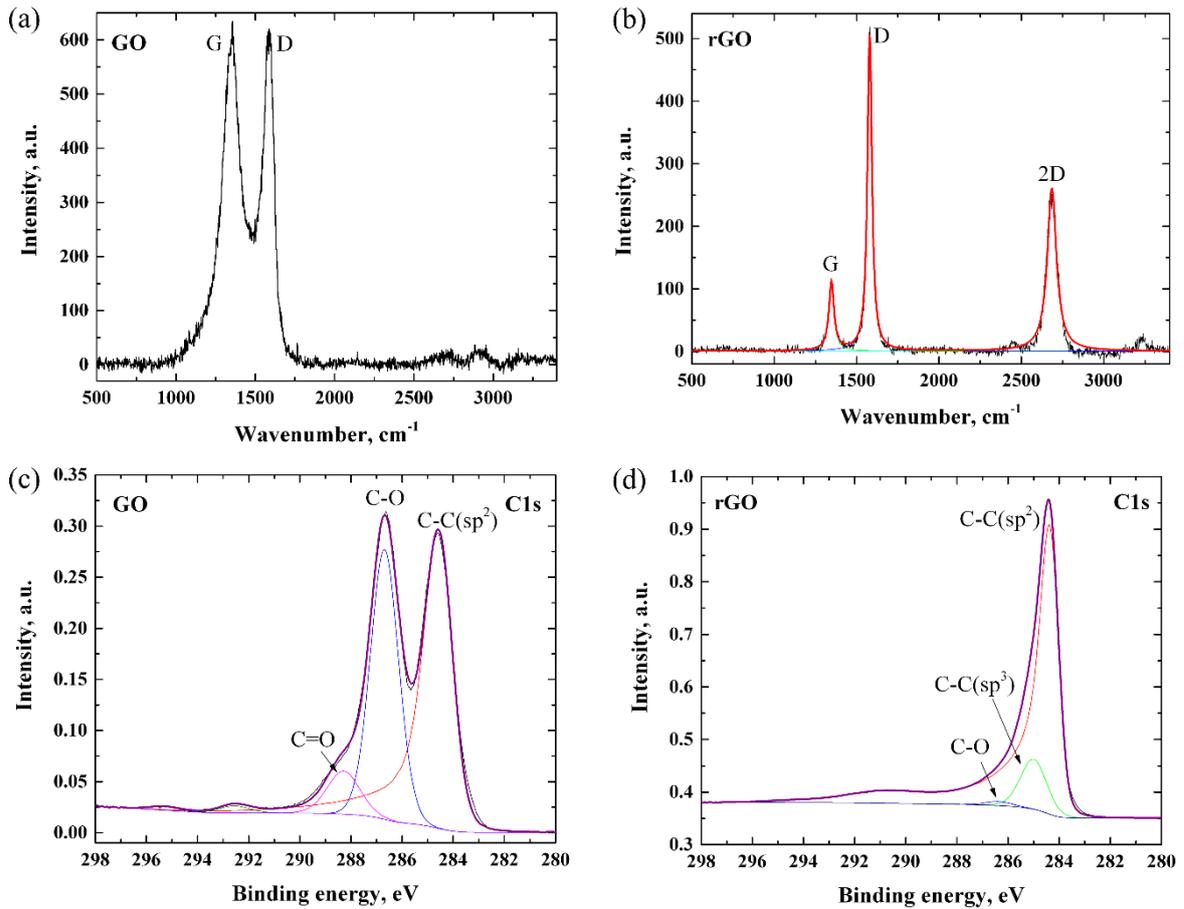

**Figure 2**. Raman and XPS spectra of GO and rGO. a) and b) Raman spectrum of GO and rGO, c) and d) XPS spectrum of GO and rGO.

The obtained samples were used to measure the gas sensitivity. To evaluate the influence of the contacts, we checked the I-V characteristics of the sensor. The I-V exemplary curves are shown in Figure 3a. The sensor exhibited a linear relationship between I and V both in air and in the air with $NO_2$, which indicated the absence of significant potential barriers at the electrode interfaces and indicated that the material is characterized by Ohmic contact. Though the characteristics obtained for and rGO seem to be similar, their structural differences caused the resistance to change from 1080.3±54.0 to 1057.4±53.7 Ohm after the addition of 25 ppm $NO_2$ (Area Figure 1d). More details on the dependence of sheet resistance on fluence and overlapping is presented in Figure S3. The lowest value of sheet resistance achieved was 60 Ohm/□.

We studied the sensor response toward the vapors of reducing ($H_2S$ and $NH_3$) and oxidizing ($NO_2$) gases. The presence of the admixtures in the air reversibly changes the conductance of the prepared sensor (Figure 3 b-c). We calculated the response of the sensor according to

$$S = \frac{|G_{Gas} - G_{Air}|}{G_{Air}} \times 100\%, \qquad (2)$$

where $G_{Gas}$ and $G_{Air}$ are the conductance of the sensor under exposure to the gas vapors in the mixture with synthetic air and in synthetic air, respectively. The observed sensor response was especially pronounced for $NO_2$ vapors (approximately 5% at 50 ppm), while the responses are quite small for $H_2S$ and $NH_3$ (0.1% and approximately 0.04%, respectively, at 50 ppm). The defects, shape, and surface groups on the rGO influenced its behavior. Different kinds of defects are present in our structures as seen in Table S1 and Figure 1. The type of defects have a substantial impact on the properties of the material. Point defects have no impact on the resistivity, while micro defects lead to resistivity changes during molecular adsorption.[43] Defects in the graphene change the adsorption energy of molecules.[36] Some gases can act as hole or electron dopants, which can lead to increases or decreases in resistivity.[44] An oxidizing gas such as $NO_2$ can act as a *p*-dopant, increasing the conductance by injection of holes via withdrawing electrons (and hole promotion) from residual epoxy and carboxyl groups naturally present in the rGO;[45] reducing gases have the opposite effect. At the same time, the mechanism also must involve penetration of the gases through the layers of the rGO where the water and analyte molecules can facilitate the change in conductance by interacting with the functional groups.[44]

The responses of the sensor toward $NO_2$ at 25 ppm in mixtures with air of different humidities are presented in Figure 3b. The increase in the response caused by the increase in humidity might be due to presence of water molecules, which affect the conductivity of the material.[46] The tested analytes are easily dissolved in water (unlike some poorly soluble organic molecules), which might also facilitate the observed increase in the response as shown in the exemplary case.

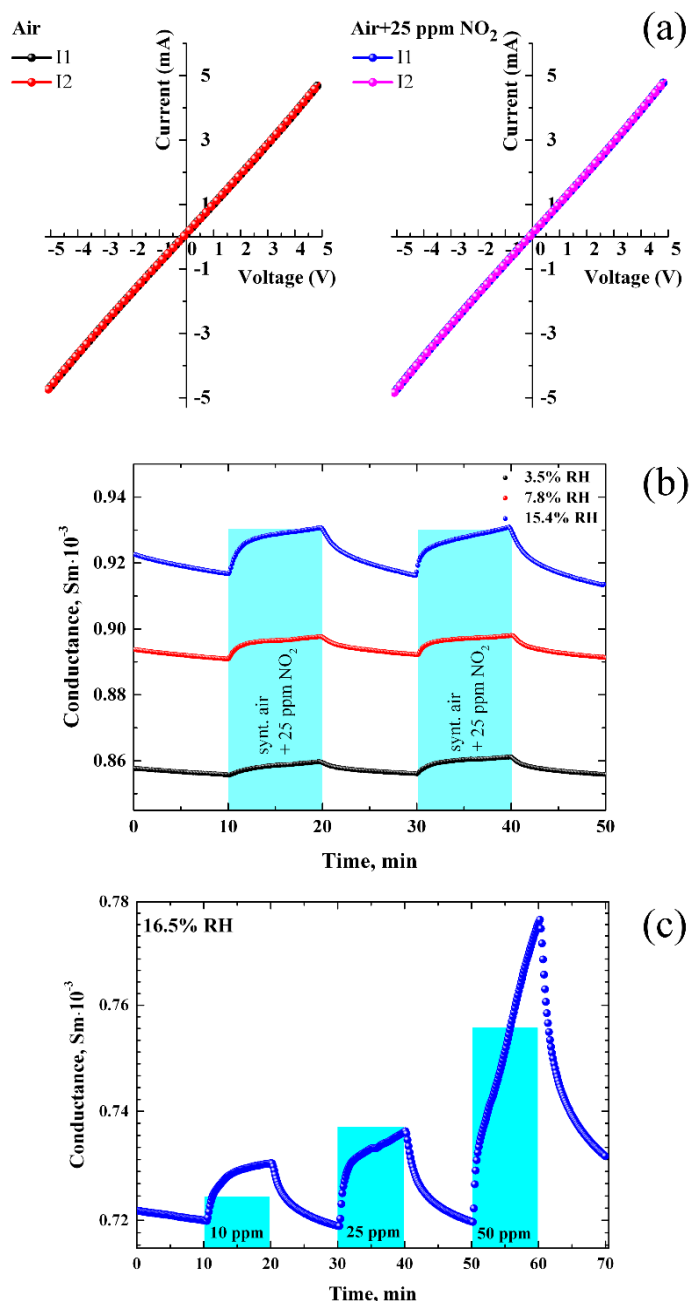

**Figure 3**. Evaluation of the gas-sensing properties of the rGO sensor at 25 °C. a) I-V characteristics of the rGO sensor in synthetic air (left) and in synthetic air mixed with 25 ppm $NO_2$ (right); I1 and I2 showed opposite directions of scanning. b) Response of the rGO sensor toward 25 ppm $NO_2$ mixed with synthetic air at 3.5, 7.8 and 15.4% relative humidity (RH). c) Response of the rGO sensor toward $NO_2$ vapors in a mixture with air at 16.5% RH at various vapor concentrations (10, 25, and 50 ppm).

Increasing the vapor concentration favors decreases ($H_2S$ and $NH_3$, see Supplementary Materials) or increases ($NO_2$) in the conductance of the sensor. In particular, for $NO_2$, we might observe a change from 1.5% to 2.2% and 5% corresponding to changes from 10 to 25 and 50 ppm of the analyte in the mixture with synthetic air (15.8% RH) as shown in Figure 3c. A dependence on concentration was also observed for H2S and NH3; however, they were not as pronounced.

## 3. Discussion

To estimate the film temperature at the focal point of the laser at the initial moment of laser treatment, we used the heat transfer equation for T(t,r,z) in cylindrical coordinates to the volume of the treated region ($0 \leq r \leq D/2$, $0 \leq z \leq H$):

$$\rho \times c \times \frac{dT}{dt} = Q(z) + \frac{d(\lambda \frac{dT}{dz})}{dz} + \frac{d(r\lambda \frac{dT}{dr})}{r}, \quad (3)$$

where D is the diameter of the laser beam; H is the film thickness; $\rho$, c and $\lambda$ are the local density, thermal capacity and thermal conductivity of the film, respectively; and Q(z) is the power density of the laser irradiation along the depth of the film (along the z-axis).

Initially, the values were $\rho \approx 1.8$ g cm$^{-3}$, $c \approx 0.71$ J g$^{-1}$ K$^{-1}$ and $\lambda \approx 0.02$-$0.04$ W cm$^{-1}$ K$^{-1}$,[47] and after laser reduction, a significant morphological transformation occurs; thus, along with the changes in absorption,[48] amount of sp$^2$ – hybridized bonds increase,[28] and as a consequence, the heat transfer coefficient changes ($\lambda \approx 0.4$ W cm$^{-1}$ K$^{-1}$,[47] $\lambda \approx 1$ W cm$^{-1}$ K$^{-1}$).[28] Due to the significant change in the parameters of the structural properties due to laser reduction, it is difficult to solve equation 3. Therefore, the temperature for the first several laser pulses must be estimated.

The following experimental parameters were used to obtain most of the results in this work: average irradiation intensity $I(z=0)=7.56 \times 10^4$ W cm$^{-2}$, peak intensity $I_{peak}(z=0)=4.8 \times 10^5$ W cm$^{-2}$ in the pulse, $t_{pulse}=200$ ns, with a pause time of $t_{pause}=800$ ns and $t_{period}=1000$ ns. If we integrate equation 3 and disregard the heat transfer, we will obtain the value of temperature for the first laser pulses. Laser intensity distribution through the depth of the film obeys the Beer–Lambert law. The absorption coefficient of GO measured for the laser wavelength was $\gamma \approx 3000$ cm$^{-1}$.

$$I(z) = I(z=0) \times e^{-\gamma \times z} \quad .$$

(4)

The equation for the power density of the absorbed laser radiation is as follows:

$$Q(z) = -\frac{dI}{dz} = \gamma \times I(z).$$

(5)

From equation 3, we obtain the following values of increase in temperature, ΔT(z), for one laser pulse:

$$\Delta T(z) \approx \frac{t_{pulse} \times \gamma \times I_{peak}(z)}{\rho \times c} = 226\ K. \tag{6}$$

For the near-surface region, heat addition is ΔT(z=0) ≈ 226 K, and due to the low initial heat transfer coefficients of the GO film, the heated regions will not have enough time to cool between pulses, and therefore, after the next pulse, the material will be heated to a temperature of T(z=0)>700 K, which will lead to an active state of GO reduction in the heated region with a corresponding increase of λ.[28]

To estimate the maximum temperature, we considered the case of cylindrical symmetry for a region with a radius of R=D/2 under the laser beam. As a temperature increases, T(r<R) in the considered region, the neighboring regions will start to heat with a corresponding increase in the heat transfer coefficient, λ(T). For the model case, we will assume power-law dependence of coefficient λ(T)= $\lambda_0 T^a$ with its initial value at 300 K of λ(T=300 K)= λ(GO)≈0.02 W cm$^{-1}$ K$^{-1}$; in addition, the maximal value of λ(T) should not exceed the value of λ (rGO) ≈ 1 W cm$^{-1}$ K$^{-1}$. In this case, the quasi-stationary radial temperature profile of the film outside the laser beam can be described with the following equation: d(rλdT/dr)/dr=0. Integration of this expression in each layer over *r* leads to equation 7.

$$\lambda_0 \times T^a \times \frac{dT}{dr} = -\frac{C}{r}. \tag{7}$$

The integration constant in equation 7 is determined from the boundary conditions at the edge of the heated region (r=R, absorbed power of P=γI(z)πR² in the region r<R is mostly removed via radial heat transfer): $2\pi C(z)=\gamma I(z)\pi R^2$. Another boundary condition used was T(z, $R_{ext}$)=300 K on the outer border of $R_{ext}$ where the temperature is close to the initial value. Considering this condition, we integrate equation 6 and obtain radial profiles.

$$T(z,r) = (300^{a+1} - (C(z) \times \frac{a+1}{\lambda_0} \times \ln\frac{r}{R_{ext}}))^{\frac{1}{a+1}}. \tag{8}$$

The maximum achievable temperature, $T_{max}(z,r=0)$, at the center of the laser beam will exceed T(z,R); however, due to the high coefficient of heat transfer, $\lambda \approx \lambda(rGO)$, overheating in the region of the laser spot will not be critical (the radial temperature gradient estimation provided values of not more than 200K). The influence of the choice of $R_{ext}$ on the analytical profiles (equation 7) for z=0 are demonstrated in **Figure 4**a at a=2, I(z=0)=7.56×10$^4$ W cm$^{-2}$ and three values of $R_{ext}$ (30, 45 and 60 μm).

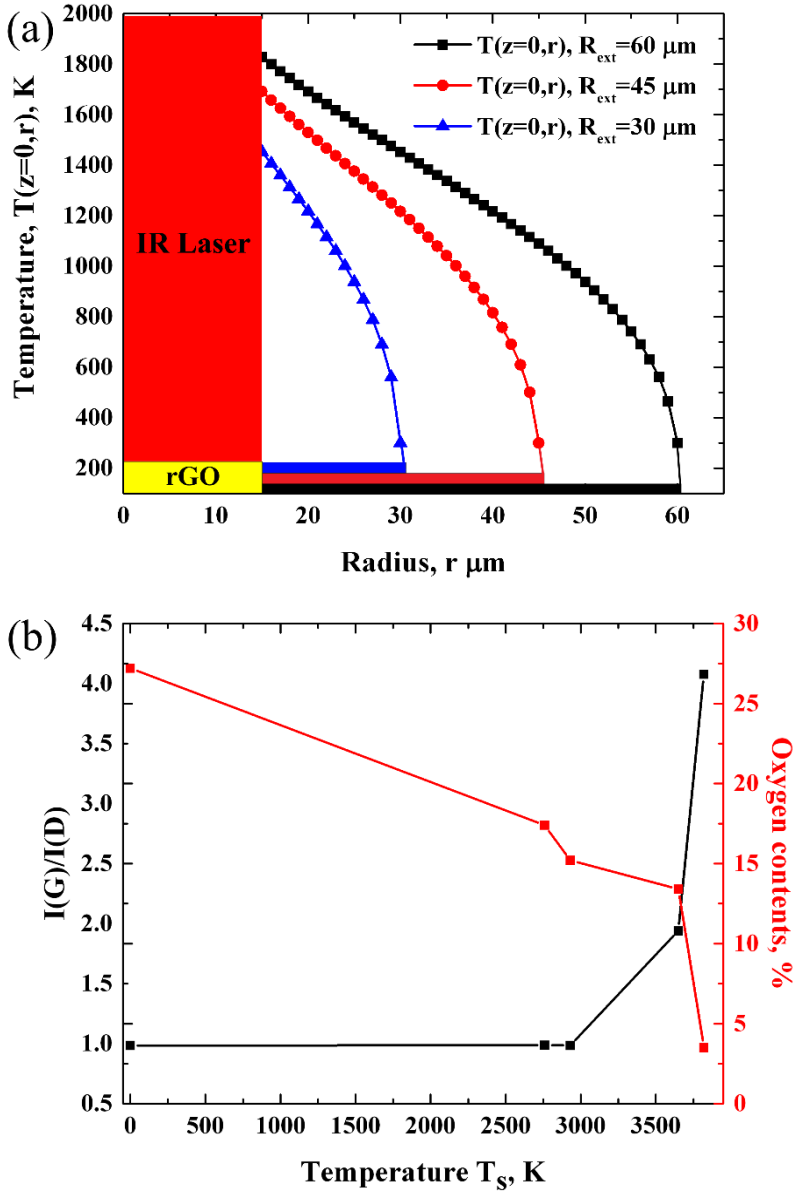

**Figure 4.** a) Temperature distribution in the GO according to equation 7 for laser spot a=2, I(z=0)=9.62×10⁴ W cm⁻² and $R_{ext}$=30, 45 and 60 μm. b) Peak I(G)/I(D) ratio and remaining oxygen fraction as a function of temperature, $T_s$; using the Planck black-body radiation formula provided the best approximation of the measured spectral background curves.

The temperature of the film did not exceed 2000 K, and thus, the heat transfer coefficients calculated for a=2 are less than λ(T=2000 K)=0.02×(2000/300)ᵃ=0.9 W cm⁻¹ K⁻¹, which was proved to be less than λ for rGO.

The case observed above is a simplified model, and real laser heating of the film would be more complicated. Due to the changes in morphology, the absorption of the film will also change, leading to an increase in temperature. Morphological changes will also occur deeper inside the film, thus simultaneously significantly reducing the density of the material and increasing the absorption coefficient. These changes should lead to a significant temperature increase in the carbon structures, especially at the surface tips, which are the further from the main material of the film. These spots should have higher temperatures due to the great heat transfer from the absorbed laser power along the layers of the rGO. In addition, as we have shown in the model for the initial heating of the GO film, the typical laser reduction process with constant displacement of the laser spot on the surface of the film will occur not in the initial GO structures but in the preheated and thus partially reduced regions of the film. This observation may play a significant part in the determination of the differences between the GO films and GO reduced with laser spots with various displacement velocities.

The described reduction parameters were studied via spectroscopy methods, and the background curves were approximated by Planck's radiation formula with the appropriate temperatures, $T_s$. The quality of the rGO is dependent on the temperature, and an increase in temperature ($T_s$ to ~3800 K) improves the quality of the rGO, which was confirmed by the XPS and Raman studies (Figure 4b). The power per square centimeter in our method is much higher than the ablation threshold, which is consistent with what was reported previously.[24] In our methods, the flow of the hot sputtered material (even plasma at high values of power per square cm) interacts with the surrounding atmosphere, cools, and participates in various chemical reactions. In particular, carbon atoms and clusters will be actively involved in oxidation processes, such as the fast reaction of $C + O_2 <=> CO + O$ (rate constant of $k \sim 10^{-11}$ cm$^3$ s$^{-1}$) with following reaction: $CO + O + M <=> CO_2 + M$ (in our case, the third species (M) scan be $N_2$ or $O_2$).[49]

## 4. Conclusions

In summary, we introduced an up-scalable, easily controllable and fast laser method to fabricate high quality reduced graphene oxide. The process does not require vacuum equipment or chemical reagents and can be performed at ambient air conditions. The method allows us to reduce 25 cm$^2$ GO film during 1 min. The material contains less than 3% of oxygen and have an average crystalline size of ~90 nm, which according to our best knowledge is the best result obtained for laser graphene oxide reduction. Our theoretical studies predicted that the temperature of the reaction can reach several thousands of Kelvin during the reduction process, which allowed us to improve the quality of the produced rGO. In addition, for the first time we show that obtained material can be used a gas high sensitivity sensor with a detection limit down to 10 ppm.

**4. Experimental Section**

*4.1 Materials*

Graphite (Timcal, Timrex KS 15) was used for the preparation of the GO. A standard Hummers procedure was performed for the graphite oxidation[3]. The obtained mixture was centrifuged and carefully washed with deionized water until the pH of the water was 5. Finally, 5 g L$^{-1}$. GO solution was prepared. A drop casting technique was used to cover the glass or silicon substrate. The obtained films were dried under ambient conditions.

*4.2. Reduction technique*

Laser microstructuring was performed by a fiber pulse laser IRE- Polus YLPM-1-4x200-20-20 with a wavelength of 1.064 μm. The laser was used in single TEM00 mode, and at a beam quality factor of M$^2$=1.005. The laser pulse duration was varied from 4 to 200 ns, and the pulse frequency was varied from 1.6 to 1000 kHz. The beam was focused to a spot 30 μm in diameter. The focusing point position can be changed by a scanning Galvo mirrors system with a variable speed of up to 15 m/s along the surface of the sample. Laser processing was performed in the shape of lines with a density of 20 lines per mm. The relationship between the pulse frequency and the speed of the focal point movement defines the pulse overlapping, which was varied from 1 to 3000. The total laser fluence was varied from 0.1 to 100 J cm$^{-2}$.

*4.3. Structural analysis*

The morphologies of the samples were analyzed using SEM (Carl Zeiss Supra 40). The structures of the samples were investigated by Raman spectroscopy, which was performed using a Jobin

Yvon instrument (LabRAM HR800 UV−visible−NIR), Thermo Scientific DXRxi with an excitation wavelength of 532 nm and at 20% percent of the maximum laser power. Fourier transform infrared spectroscopy (FTIR) data were recorded by using a Bruker Vertex V70. XPS were recorded by an Axis Ultra DLD (Kratos) spectrometer with monochromatic Al Kα radiation using an analyzer energy transmission of 20 eV. Temperature estimation during laser reduction was performed by measuring the black-body spectrum using an Avesta ASP-100 spectrometer. The obtained spectra were approximated by Planck's radiation formula.

*2.4 Gas sensor measurements*

To study gas characteristics, Au contacts were deposited on the rGO films. Deposition was performed by magnetron sputtering of high purity Au (99.999%) in an argon atmosphere (99.999% purity). The rate of Au deposition in this process was 1 nm s$^{-1}$. The distance between electrodes was 2 mm. For the gas interaction study, samples were mounted in a custom experimental setup (see Figure S5). The sensor has been installed on a temperature-controlled stage with electrical connections (Linkam THMS350EV equipped with LNP96 liquid nitrogen cooling system) where its conductance when subjected to different gas environments could be monitored. The conductance was measured by a Keysight 34972A LXI Data Acquisition/Switch Unit equipped with a 34901A 20-channel Armature Multiplexer and a 34907A Multi-Function Module. The measurements were conducted in the constant flow mode at a flow rate of 100 sccm. The experiments were conducted at 25 °C, which was controlled by Linkam unit. The relative humidity, RH, was controlled by a Testo 645 sensor. Prior to each measurement, the sensor was annealed at 120 °C for 15 min in the synthetic air with a set humidity level that had been purified to remove adsorbed species.[34,50]

**Acknowledgments**

This work was supported by Russian Science Foundation (grant #17-72-10307).

# Supporting Information

## Ambient condition production of high quality reduced graphene oxide

*Sergey E. Svyakhovskiy\*, Stanislav S. Evlashin, Fedor S. Fedorov, Yuri A. Mankelevich, Pavel V. Dyakonov, Nikita V. Minaev, Sarkis A. Dagesyan, Konstantin I. Maslakov, Roman A. Khmelnitsky, Nikolay V. Suetin, Iskander Sh. Akhatov and Albert G. Nasibulin*

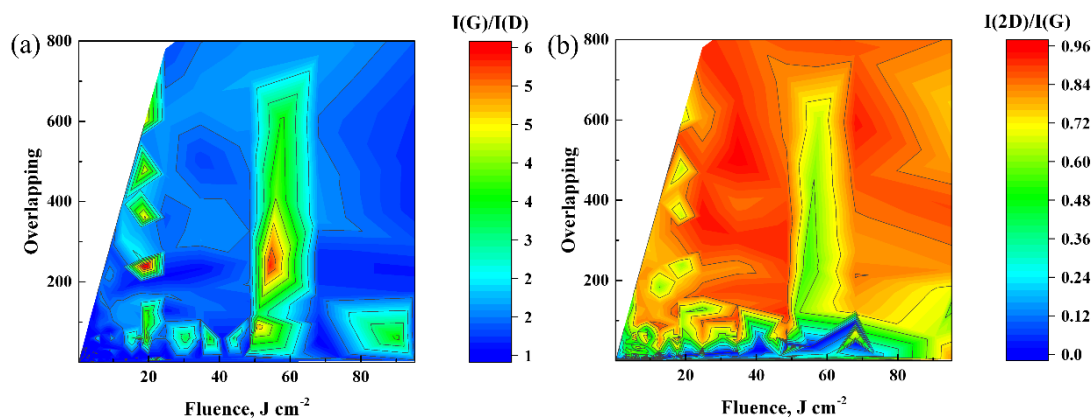

**Figure S1.** a) Map for Raman peak I(G)/I(D) ratio at different fluences and degrees of overlap. b) Map for Raman peak I(2D)/I(G) ratio at different fluences and degrees of overlap.

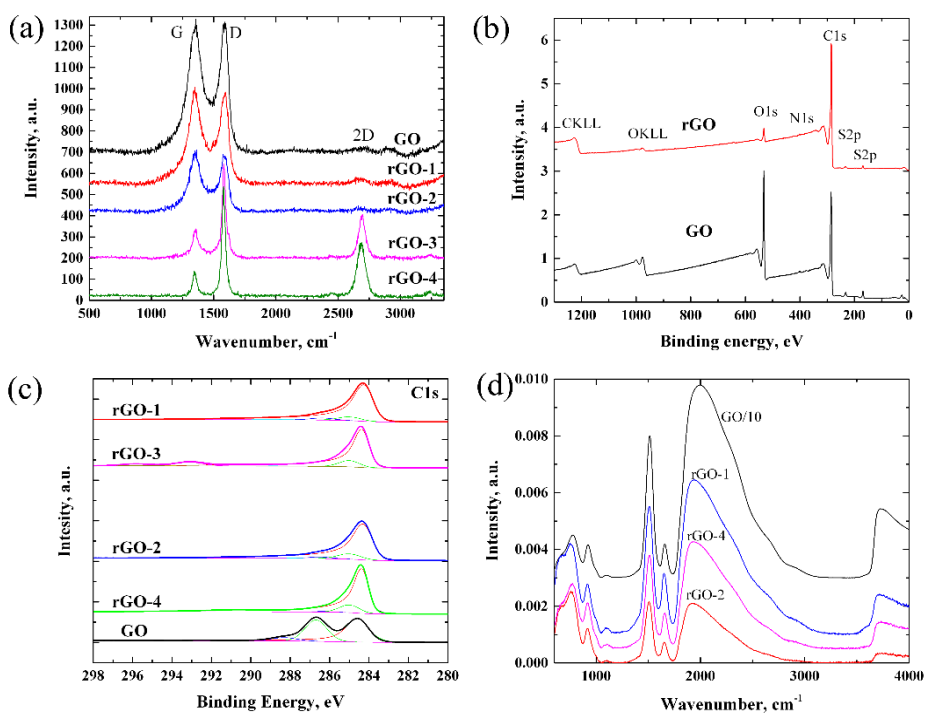

**Figure S2.** a) Raman spectra for the different sets of reduction conditions. b) XPS spectra of GO and rGO. c) XPS C1s for the different sets of reduction conditions. d) FTIR spectra for different sets of reduction conditions.

**Table 1.** Concentrations of elements in the studied samples (at.%) calculated based on the high-resolution XPS spectra.

| Sample | O | N | C | K | S |
|---|---|---|---|---|---|
| GO | 27.2 | 0.6 | 70.5 | 0.2 | 1.5 |
| rGO1 | 17.4 | 3.1 | 76.2 | – | 3.3 |
| rGO2 | 15.2 | 2.6 | 79.5 | – | 2.7 |
| rGO3 | 13.4 | – | 82.1 | 2.1 | 2.4 |
| rGO4 | 3.5 | 0.5 | 95.3 | – | 0.7 |

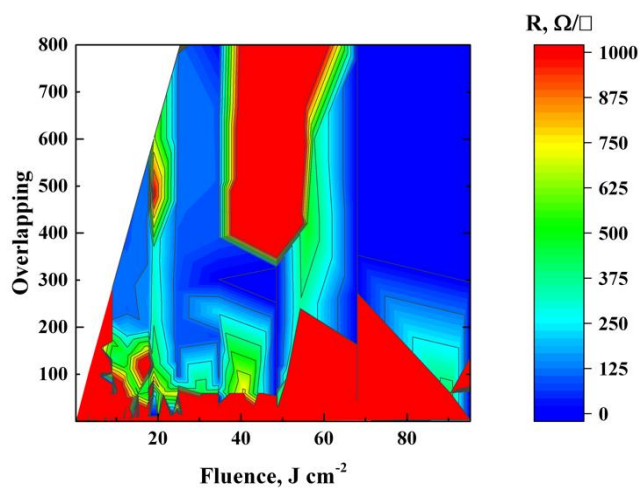

**Figure S3.** Resistance of the different reduction areas.

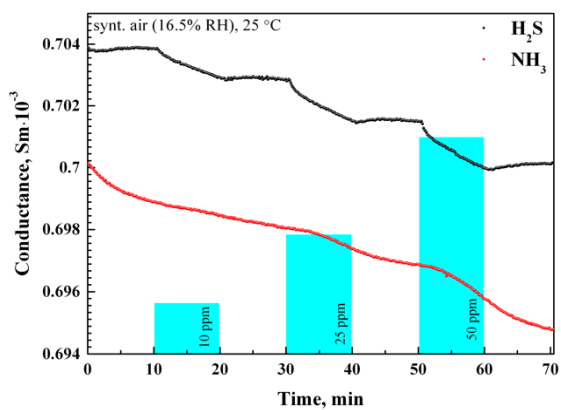

**Figure S4.** The response of the rGO sensor to NH$_3$ and H$_2$S vapors mixed with synthetic air at 25 °C

Figure S4 shows the response of the rGO sensor to H$_2$S and NH$_3$ vapors mixed with synthetic air at 25 °C. The responses are minor and are equal to 0.1% and approximately 0.04% for 50 ppm of H$_2$S and NH$_3$ vapors, respectively, mixed with synthetic air (Figure S1a).

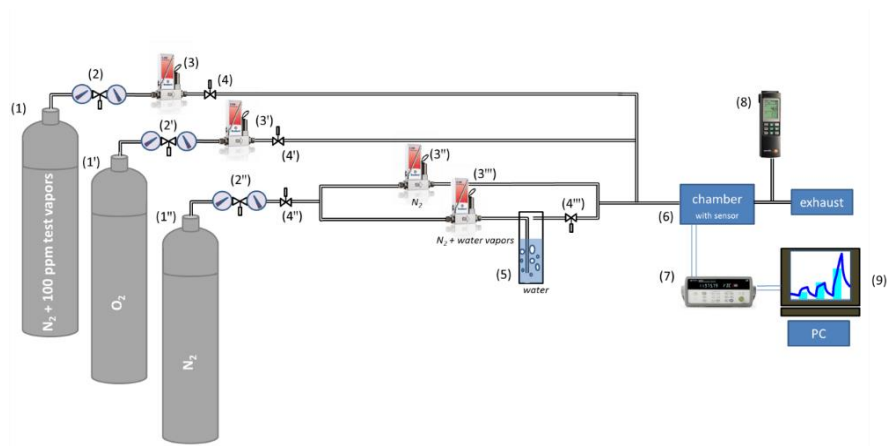

**Figure S5.** Scheme of the experimental setup: (1-1'') gas tanks; (2-2'') gas pressure valves, (3-3''') mass flowmeters; (4-4''') valves; (5) bubbler filled with deionized water; (6) chamber; (7) data acquisition/switch unit; (8) humidity sensor; and (9) PC terminal.

The experimental setup is presented in Figure S5. Oxygen and nitrogen were mixed in the proper proportion to produce the synthetic air. The humidity level was adjusted by purging with nitrogen using deionized water in a bubbler, and that process was monitored by a Testo 645 sensor. The gas tanks were connected to the pipeline via valves. The operation of the flowmeters was controlled by a PC running @LabVIEW software. The gas pipeline was connected to a chamber with the rGO sensor. The exhaust gases were released to a fume hood. The conductance change was recorded by a Keysight 34972A LXI Data Acquisition / Switch Unit. The temperature was maintained at 25 °C using a Linkam THMS350EV.